\documentclass[aps,pre,twocolumn,superscriptaddress]{revtex4}
\usepackage[usenames,dvipsnames]{color}

\usepackage{graphicx,color}
\usepackage{bm}
\usepackage{amsmath}
\usepackage{verbatim}
\usepackage{amssymb}

\begin{document}
\title{Telling time with an intrinsically noisy clock}
\author{Andrew Mugler}
\affiliation{Department of Physics, Columbia University, New York, NY 10027}
\author{Aleksandra M. Walczak}
\affiliation{Princeton Center for Theoretical Science, Princeton University, Princeton, NJ 08544}
\author{Chris H. Wiggins}
\affiliation{Department of Applied Physics and Applied Mathematics, Center for Computational Biology and Bioinformatics, Columbia University, New York, NY 10027}

\date{\today}

\begin{abstract}
Intracellular transmission of information via chemical and transcriptional networks is thwarted by a physical limitation: the finite copy number of the constituent chemical species 
introduces unavoidable intrinsic noise. 
Here we provide a method for solving for the complete probabilistic description of intrinsically noisy oscillatory driving. We derive and numerically verify a number of simple scaling laws. Unlike in the case of measuring a static quantity, response to an oscillatory driving can exhibit a resonant frequency which maximizes information transmission. Further, we show that the optimal regulatory design is dependent on the biophysical constraints (i.e., the allowed copy number and response time).
The resulting phase diagram illustrates under what conditions threshold regulation outperforms linear regulation.
\end{abstract}

\maketitle

\newcommand{\bra}[1]{\langle{#1}|}
\newcommand{\ket}[1]{|{#1}\rangle{}}
\newcommand{\avg}[1]{\langle{#1}\rangle}
\newcommand{\bracket}[2]{\langle{#1}|{#2}\rangle{}}
\newcommand{\beqn}{\begin{eqnarray}}
\newcommand{\eeqn}{\end{eqnarray}}
\newcommand{\beq}{\begin{equation}}
\newcommand{\eeq}{\end{equation}}
\newcommand{\abs}[1]{|#1|}
\newcommand{\A}{{a^{\dagger}}}
\newcommand{\av}[1]{\langle{#1}\rangle{}}
\newcommand{\mk}[2]{\newcommand{#1}{#2}}
\newcommand{\rmk}[2]{\renewcommand{#1}{#2}}
\newcommand{\ip}[2]{\langle#1|#2\rangle}
\mk{\bbn}{b^+_nb^-_n}
\mk{\bbm}{b^+_mb^-_m}
\mk{\bbnn}{b^+_nb^-_n(n)}
\mk{\bbmn}{b_m^+b_m^-(n)}
\mk{\bb}{\bd_n\bl_n}
\mk{\bbbn}{\bd_n{\bar{b}^-_n}}
\mk{\bbbm}{\bd_m{\bar{b}^-_m}}
\mk{\bmb}{{\bar{\bl_m}}}
\mk{\bG}{{\bf{{G}}}}
\mk{\bmn}{\hat{b}_m^-(n)}
\mk{\bnn}{\hat{b}_n^-(n)}
\mk{\bbarn}{\hat{\bar{b}}_n^-}
\mk{\bbarm}{\hat{\bar{b}}_m^-}
\mk{\betan}{\hat{\beta}^-_n}
\mk{\betam}{\hat{\beta}^-_m}
\mk{\bbar}{\bar{b}^-}
\mk{\bd}{\hat{b}^+}
\mk{\ad}{\hat{a}^+}
\mk{\bl}{\hat{b}^-}
\mk{\al}{\hat{a}^-}
\mk{\qn}{q(n)}
\mk{\gn}{g(n)}
\mk{\qh}{\hat{q}_n}
\mk{\gh}{\hat{g}_n}
\mk{\qb}{\bar{q}}
\mk{\gb}{\bar{g}}
\mk{\Dn}{\hat{\Delta}_n}
\mk{\Gah}{\hat{\Gamma}_n}
\mk{\Lah}{\hat{\Lambda}_n}
\mk{\Deh}{\hat{\Delta}_n}
\mk{\Gjk}{G^{jk}}
\mk{\bj}{\<j|}
\mk{\bk}{\<k|}
\mk{\Gl}{{G_\ell}}
\mk{\Del}{\Delta}
\newcommand{\e}[1]{{\rm{e}}^{#1}}
\mk{\ehat}{\bf{\hat{ e}}}
\rmk{\L}{{\cal L}}
\rmk{\H}{\hat{H}}
\mk{\Om}{\hat{\Omega}}
\mk{\n}{\langle n\rangle}

It has long been recognized \cite{berg} that the ability
to measure biochemical quantities, e.g., concentrations, is
intrinsically thwarted by the small copy numbers present
at the scale of the cell. This observation has launched
considerable experimental investigations as to how high-fidelity
signal transmission can occur within single cells \cite{mettetal2008frequency, repressilator}, along with an associated literature in mathematical and computational techniques
for modeling such noisy information transmission \cite{tostevin2009mutual, TanaseNicola, Tkacik2}. From the
perspective of biological design -- either to understand the
mechanisms which lead to observed biology or to create synthetic
systems with desirable properties -- these works investigate how
regulatory elements which comprise biological systems
function in the presence of intrinsic noise \cite{bialek2005physical}.

\begin{figure}
\begin{center}
\includegraphics[scale=0.47]{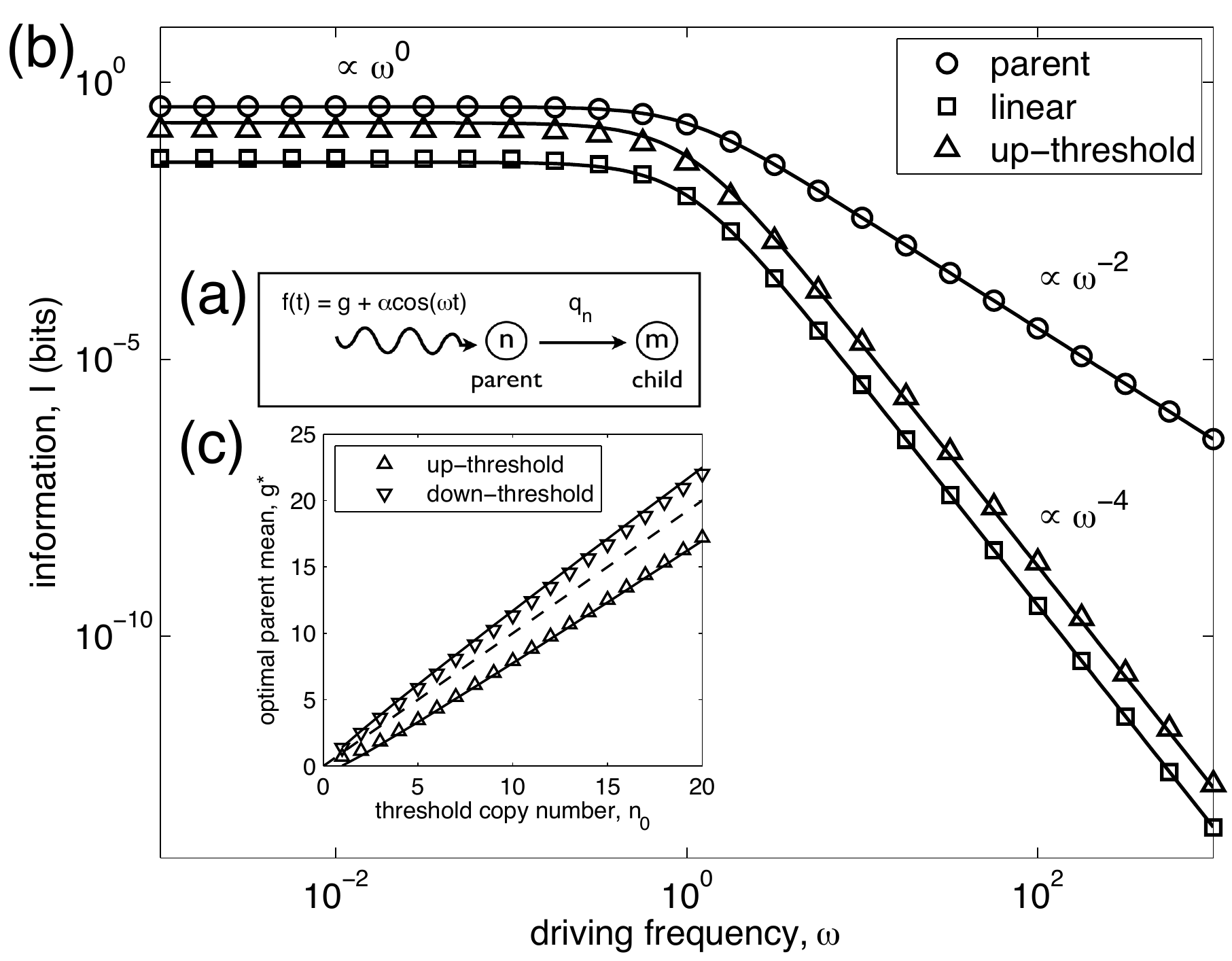}
\linespread{1}
\caption{(a) A transcription factor (the `parent') with copy number $n$ 
is driven by an oscillatory creation rate $f(t)=g+\alpha\cos\omega t$ and regulates via $q_n$ the expression of a second species (the `child') with copy number $m$.
(b, c) Numerical confirmations (data points) of analytic expressions (lines) derived in the small-information limit: Eqns.\  \ref{Ipar2} (circles)
and \ref{Im} top line (squares) and bottom line (triangles)
in (b), and Eqn.\ \ref{gstar} for both up- (up-triangles) and down-regulation (down-triangles) in (c).  Parameters are $g=n_0=1$ and $c=0.1$ for (b), $\omega=1$ for (c), and $\alpha=\rho=\Delta=1$ and $q_0=0$ for both, yielding small parameters $|\nu_1|\le0.5$ (Eqn.\ \ref{nu1}), $|\mu_1|\le0.05$ (linear; \ Eqn.\ \ref{mu1}) and $|\mu_1|\le0.184$ (threshold; \ Eqn.\ \ref{mu1}).}
\label{smallcheck}
\end{center}
\end{figure}

In earlier work we showed how the `spectral method' leads to
an efficient and accurate numerical technique, which permits optimization to
reveal the information-optimal design of a transcriptional cascade
in the presence of intrinsic noise in the statistical steady-state \cite{mtv,cspan}. 
We here turn our attention to 
the simplest model of the dynamic case, illustrated in Fig.\ \ref{smallcheck}(a),
in which a single transcription factor (the `parent') with copy number $n$ 
is driven by an oscillatory creation rate $f(t)=g+\alpha\cos\omega t$ and regulates the expression of a second species (the `child') with copy number $m$; the regulation is modeled via the child's creation rate $q_n$.
This model captures the noisy downstream response to oscillation, e.g., 
the cell cycle, without limiting the results to a particular mechanism
for generating oscillations (e.g., via cell division \cite{tyson}, 
repressive cycles \cite{repressilator},
or 
activation-repression circuits \cite{hastyreview}).
We show how the optimal design -- i.e., the choice of 
linear-vs.-cooperative and up-vs.-down regulation --
is determined by the physical demands in terms
of allowed copy number and response time. Further, while our intuition
from understanding how best to measure static signals suggests that 
slower response time is always more accurate \cite{berg}, we illustrate
how oscillatory driving leads to an information-optimal driving frequency,
and compute how this frequency depends on copy number.

In the spectral method \cite{mtv} we exploit the linearity of the master equation
${\dot p}_{nm}=-\left({\cal L}_n[f(t)]+\rho{\cal L}_m[q_n]\right)p_{nm}$, the
equation of motion for the joint probability of observing $n$ and $m$ copies of
the parent and child, respectively,
and expand its solution in terms of the natural eigenfunctions of the 
birth-death process with constant creation and decay.
For a birth-death process expressed in terms of an arbitrary creation rate $h$ on species $s$ the positive semidefinite operator ${\cal L}_s$ acts as ${\cal L}_s[h] p_s=sp_s-(s+1)p_{s+1}+hp_s-hp_{s-1}$;
time is normalized via the parent decay rate (in these units $\rho$ is the child decay rate).
To study dynamics we 
also
Fourier transform
in harmonics of the driving frequency $\omega$, 
\beq
{p}_{nm}(t) = \sum_{j=0}^\infty\sum_{k=0}^\infty\sum_{z=-\infty}^\infty p_{jk}^z \ip{n}{j} \ip{m}{k} {\rm e}^{-iz\omega t},
\label{gendef}
\eeq
where the parent and child 
eigenfunctions
(or `spectral modes')
enjoy ${\cal L}_n[g]\ip{n}{j}=j\ip{n}{j}$ and ${\cal L}_m[{\bar q}]
\ip{m}{k}=k\ip{m}{k}$, respectively.
Just as in \cite{cspan,mtv}, we introduce a gauge ${\bar q}$ to define the basis; the analytic results below are independent of this choice.

The master equation then becomes an 
algebraic relation among
the expansion coefficients $p_{jk}^z$:
\beq
-i\omega z p_{jk}^z = -(j+\rho k)p_{jk}^z + \frac{\alpha}{2} \sum_\pm p_{j-1,k}^{z\pm1}
	- \rho\sum_{j'}\Delta_{jj'}p_{j',k-1}^z,
\label{eom}	
\eeq
where $\Delta_{jj'} = \sum_n \ip{j}{n} (\qb - q_n) \ip{n}{j'}$. 
Algorithmically we
(i)
initialize with 
$G_{00}^z = \delta_{z0}$ (set by normalization),
(ii) exploit the subdiagonality in $k$, and 
(iii) for each $k$, exploit the subdiagonality in $j$;
no matrices need be inverted.

Efficient computation of $p_{nm}(t)$ allows optimization of the mutual information 
$I(\phi,n)$
between the input variable---the phase $\phi = \omega t$ of the driving oscillation---and the output variable---the copy number of either the parent or the child:
\beq
\label{Inum}
I(\phi,n) = \int_{0}^{2\pi} d\phi\, \sum_{n} p(n|\phi) p(\phi) \log \frac{p(n|\phi)}{p_n^0},
\eeq
where $p(n|\phi) \equiv p_n(t)$, $p(\phi) = 1/2\pi$, and $p_n^0 = \int_0^{2\pi} d\phi\, p(n|\phi)p(\phi)$ is the time-averaged distribution \cite{Shannon}.  Eqn.\ \ref{Inum} is integrated numerically during optimization.

In parallel with the numerical efficiency afforded by the spectral method, considerable progress can be made analytically.  The dynamics of the parent, for example, can be found exactly: the equation for $p_n(t)$ (obtained by summing the master equation over $m$) is easily solved using either the method of characteristics (Sec.\ \ref{moc}) or spectral decomposition (Sec.\ \ref{sm}).
The solution is a Poisson distribution with time-dependent mean $\nu(t) = \nu_0 + 2|\nu_1|\cos(\omega t-\gamma)$, where
\beqn
\label{nu0}
\nu_0 &=& g,\\
\label{nu1}
|\nu_1| &=& \frac{\alpha}{2\sqrt{1+\omega^2}},
\eeqn
and $\gamma = \tan^{-1}\omega$.  Since the full dynamics are known, the Fourier transform coefficients $p_n^z = \int_{0}^{2\pi} d\phi\, e^{iz\phi} p(n|\phi)/(2\pi)$ are computed by expanding the exponential in $p(n|\phi)$ and identifying the modes (Sec.\ \ref{ftp}):
\beqn
\label{pnz}
p_n^z = e^{iz\gamma} \sum_j \frac{|\nu_1|^{2j+|z|}}{j! (j+|z|)!} \left\langle n \big| 2j+|z| \right\rangle.
\label{pnzparent}
\eeqn

In the limit of weak ($\alpha\ll 1$) or fast ($\omega \gg 1$) driving, an approximation for $I(\phi,n)$ may be obtained by expanding in the small parameter $|\nu_1|$.  We first express Eqn.\ \ref{Inum} in terms of the Fourier transform $p(n|\phi) = \sum_zp_n^ze^{-iz\phi}$:
\beq
\label{IFourier}
I(\phi,n) = \sum_{n,z}p_n^z\int_{0}^{2\pi} \frac{d\phi}{2\pi} e^{-iz\phi}
	\log \left(1+\sum_{z'\ne0}\frac{p_n^{z'}}{p_n^0}e^{-iz'\phi} \right).
\eeq
Then we note that for small $|\nu_1|$, Eqn.\ \ref{pnz} is dominated by the $j=0$ term, i.e.\ $p_n^z\approx e^{iz\gamma} |\nu_1|^{|z|}\left\langle n \big||z|\right\rangle/{|z|!}$.  Since this is itself small for $z\ne0$, we expand the log in Eqn.\ \ref{IFourier} as $\log(1+x) = x - x^2/2 + \dots$ for small $x$.  The first two terms in the log expansion
(Sec.\ \ref{exp})
contain the leading-order behavior in $p_n^z$ (proportional to $p_n^1p_n^{-1} = |p_n^1|^2$); employing $\int_{0}^{2\pi} d\phi\, e^{i(z-z')\phi} = 2\pi\delta_{zz'}$ one obtains
\beqn
\label{Ipar1}
I(\phi,n) &\approx& \sum_n \frac{|p_n^1|^2}{p_n^0}
\approx |\nu_1|^2 \sum_n \frac{\ip{n}{1}^2}{\ip{n}{0}}
= \frac{|\nu_1|^2}{\nu_0} \\
\label{Ipar2}
&=& \frac{\alpha^2}{4g} \frac{1}{1+\omega^2},
\eeqn
where the second to last step uses $\ip{n}{0} = e^{-g}g^n/n!$ and $\ip{n}{1} = \ip{n}{0}(n-g)/g$ \cite{cspan} to evaluate the sum.  Eqn.\ \ref{Ipar1} shows that mutual information asymptotes to the square of the amplitude of the oscillation over the mean.
Eqn.\ \ref{Ipar2} scales like $\omega^0$ at low frequency and $\omega^{-2}$ at high frequency, demonstrating that the parent acts as a low-pass filter of information; it is numerically verified
in Fig.\ \ref{smallcheck}(b).

Although the child distribution $p_m(t)$ is not analytically accessible in general, its mean $\mu(t)$ is exactly calculable: summing the master equation over both indices against $m$ and Fourier transforming yields $\mu(t) = \sum_z \mu_z e^{-iz\omega t}$, where
\beq
\label{muz}
\mu_z=\frac{1}{1-iz\omega/\rho} \sum_n q_n p_n^z.
\eeq
In the limit of weak regulation, then, (i.e.\ when $q_n$ is near constant) we may approximate $p_m(t)$ as a Poisson distribution with oscillatory mean parameterized by the first and second Fourier mode of the exact mean, i.e.\ $\mu(t) \approx \mu_0 \pm 2|\mu_1|\cos(\omega t-\theta)$
for up- (down-) regulation,
where
$\theta = {\rm phase}(\mu_1) = \tan^{-1}\omega/\rho + \tan^{-1}\omega$.
Under this approximation, as in Eqn.\ \ref{Ipar1}, the information between the phase of the driving oscillation and the copy number of the child is the oscillation amplitude squared over the mean, i.e.\ $I(\phi,m) = |\mu_1|^2/\mu_0$ for small $|\mu_1|$.

To compare the transmission properties of both non- and highly-cooperative regulation, we study both the linear function $q_n = q_0+cn$ and the threshold function $q_n = q_0+\Delta\chi(n\in\Omega_\pm)$, respectively, where $\chi$ is a characteristic function equal to $1$ when $n$ is in the set $\Omega_+=\{n>n_0\}$ (up-regulation) or $\Omega_-=\{n\le n_0\}$ (down-regulation), and $0$ otherwise.
In these cases, the mean of the child distribution oscillates about the point
\beq
\label{mu0}
\mu_0 = \sum_n q_n p_n^0 = q_0 +
\begin{cases}
	cg & {\rm linear}
		\\
	\Delta p^0_\pm & {\rm threshold.}
\end{cases}
\eeq
Here the linear result exploits the fact that the mean of the time-averaged parent distribution $p_n^0$ is $g$ (which can be seen from the relationship between distribution moments and spectral modes, Sec.\ \ref{m2m}).
In the threshold result we define $p_\pm^0 \equiv \sum_{n\in\Omega_\pm} p_n^0 = \pi_\pm \pm \sum_{j>0} |\nu_1|^{2j}\ip{n_0}{2j-1}/(j!)^2 \approx \pi_\pm$, where $\pi_\pm \equiv \sum_{n\in\Omega_\pm}\ip{n}{0}$; the second to last step exploits the result
$\sum_{n\in\Omega_\pm}\ip{n}{j} = \pm\ip{n_0}{j-1}$ for $j>0$
(Sec.\ \ref{ibp})
and the last step takes $j=0$ in the small $|\nu_1|$ limit.
The amplitude of the oscillation of the child mean is
\beq
|\mu_1|  = \frac{\sum_n q_n |p_n^1|}{\sqrt{1+(\omega/\rho)^2}}
= \frac{1}{\sqrt{1+(\omega/\rho)^2}} \times
\begin{cases}
	c|\nu_1|
		\\
	\Delta |p^1_\pm|,
\label{mu1}
\end{cases}
\eeq
where once more the linear result (top) uses the relationship between moments and modes
(Sec.\ \ref{m2m})
and in the threshold result (bottom) we define
and approximate $|p_\pm^1| \equiv \sum_{n\in\Omega_\pm} |p_n^1| = \sum_j |\nu_1|^{2j+1}\ip{n_0}{2j}/[j!(j+1)!] \approx |\nu_1|\ip{n_0}{0}$ 
(Sec.\ \ref{ibp}).
Eqns.\ \ref{mu0} and \ref{mu1} yield the following approximations for linear (top) and threshold (bottom) regulation:
\beq
\label{Im}
I(\phi,m)
\approx \frac{gI(\phi,n)}{1+(\omega/\rho)^2} \times
\begin{cases}
	c^2/(q_0+cg)
		\\
	\Delta^2\ip{n_0}{0}^2/(q_0+\Delta \pi_\pm),
\end{cases}
\eeq
where $I(\phi,n)$ is as in Eqn.\ \ref{Ipar2}.
Eqn.\ \ref{Im} shows that the child $I(\phi,m)$ is a sharper low-pass filter than the parent $I(\phi,n)$, falling off like $\omega^{-4}$ at high frequency instead of $\omega^{-2}$; it is verified numerically in Fig.\ \ref{smallcheck}(b).  We note that since $t$, $n$, and $m$ are not Markov related, i.e.\ $p(m|t)\ne\sum_np(m|n)p(n|t)$, Eqn.\ \ref{Im} is not bound by the data-processing inequality \cite{Cover}, and it is possible to have $I(\phi,m)>I(\phi,n)$ (e.g. for linear regulation with $\omega\rightarrow0$, $q_0=0$, and $c>1$), which we have confirmed numerically
(Sec.\ \ref{nonmark}).

Eqn.\
\ref{Im}
also offers analytic intuition about the optimal placement of the parent distribution with respect to a threshold regulation function.
The derivative of Eqn.\ \ref{Im} (bottom) with respect to $g$ vanishes at $g^*$,
the information-optimal mean of the parent distribution,
\beq
\label{gstar}
g^* = \frac{n_0}{1\pm\Delta\ip{n_0}{0}/[2(q_0+\Delta\pi_\pm)]}
\eeq
(recall that dependence on $g$ is contained within $\ip{n_0}{0}$ and $\pi_\pm$; Eqn.\ \ref{gstar} is transcendental and solved iteratively).
As verified in Fig.\ \ref{smallcheck}(c), Eqn.\ \ref{gstar} shows that the parent distribution is shifted below the threshold for up-regulation and above the threshold for down-regulation.  These shifts account for the ability of up-regulation to outperform down-regulation when copy number is highly constrained (see Fig.\ \ref{phase} and discussion below), an effect we observed previously \cite{mtv} when numerically optimizing steady-state information between the first and last species in a regulatory cascade.

\begin{figure}
\begin{center}
\includegraphics[scale=0.475]{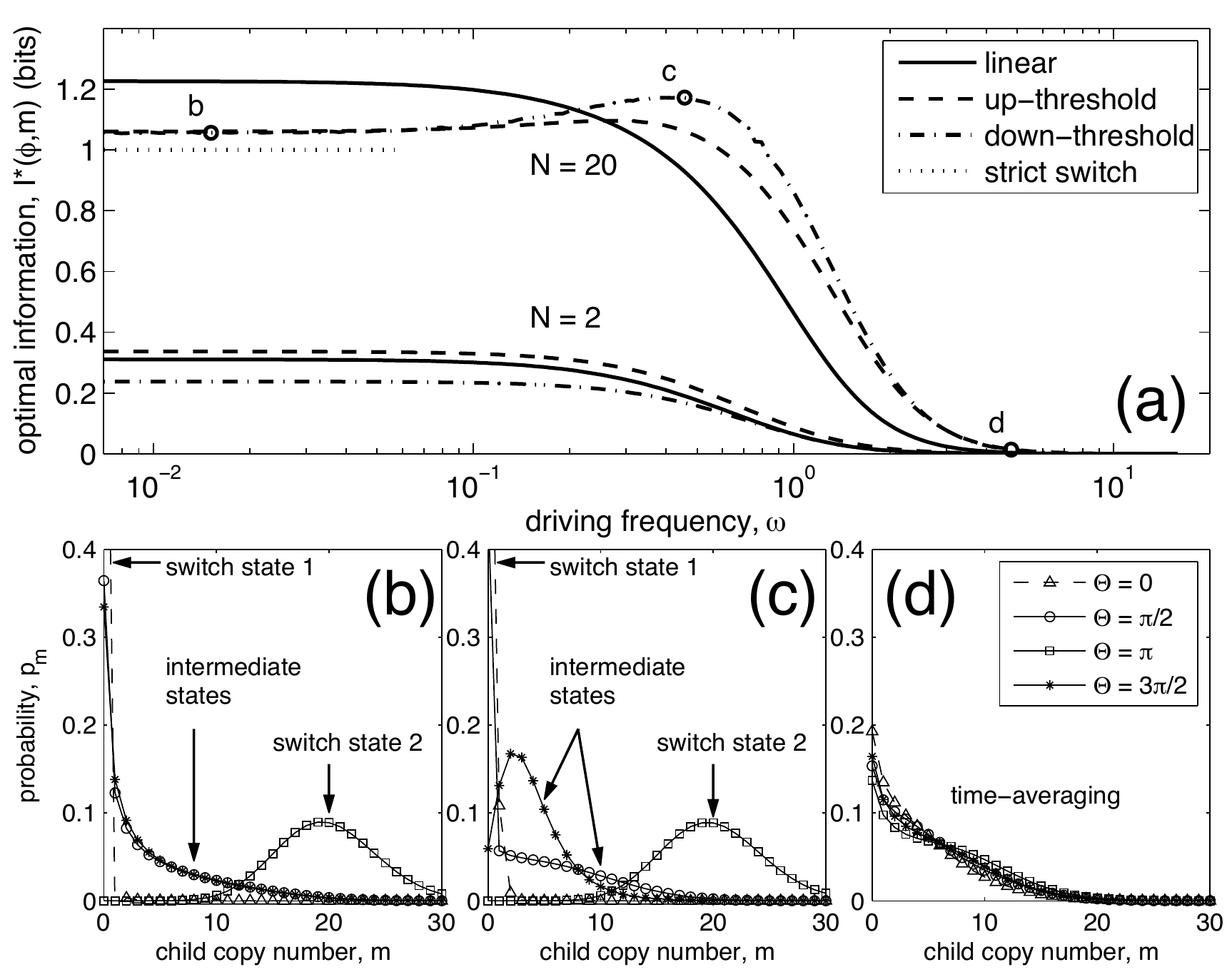}
\linespread{1}
\caption{(a) At high copy number ($N=20$, top curves) the optimal information $I^*(\phi,m)$ exhibits a resonant driving frequency (point c) for up- (dashed) and down-threshold (dot-dashed) regulation, but not for linear (solid) regulation; at low copy number ($N=2$, bottom curves), there is no resonant frequency, and slowest ($\omega\rightarrow$ 0) is best.  Panels (b-d) correspond to marked points in (a) and show the optimal child distribution for down-threshold regulation at phases $\Theta \equiv \omega t - \theta =0$, $\pi/2$, $\pi$, and $3\pi/2$ [legend in (d) applies to (b-d)]: (b) slow driving produces switch-like behavior, with long-lived high- ($\Theta=0$) and low-copy number ($\Theta=\pi$) states and brief intermediates ($\Theta=\pi/2, 3\pi/2$) in between; (c) moderate driving produces switch-like behavior with distinguishable intermediates, transmitting the most information; and (d) fast driving time-averages the parent, and thus the child, distribution.}
\label{omstar}
\end{center}
\end{figure}

The above analytic approximations offer guidance during a full numerical optimization of $I(\phi,m)$ via the spectral method.  As suggested by Eqn.\ \ref{Im}, numerical optimization confirms that $I(\phi,m)$ increases when (i) the amplitude of the driving oscillation is maximal ($\alpha=g$) and (ii) the dynamic range is maximal ($q_0=0$ and $c\rightarrow\infty$ or $\Delta\rightarrow\infty$). The slope $c$ or discontinuity $\Delta$, however, is constrained by the average copy number of the child $\mu_0$ (Eqn.\ \ref{mu0}).  Therefore for a fixed driving frequency
and fixed total average copy number $N = \avg{n}+\avg{m} = g + \mu_0$, we optimize over the single parameter $g$
by setting $\alpha=g$, $q_0=0$, and $c = \mu_0/g = (N-g)/g$ or $\Delta = \mu_0/p^0_\pm = (N-g)/p_\pm^0$; additionally we set $\rho=1$ for equal decay rates (as is typical when decay rates are dominated by cell division \cite{mtv}).
For threshold regulation, an optimization over $g$ is done at each of a set of values of the (discrete) parameter $n_0$, and the global optimum is selected.

\begin{figure}
\begin{center}
\includegraphics[scale=0.475]{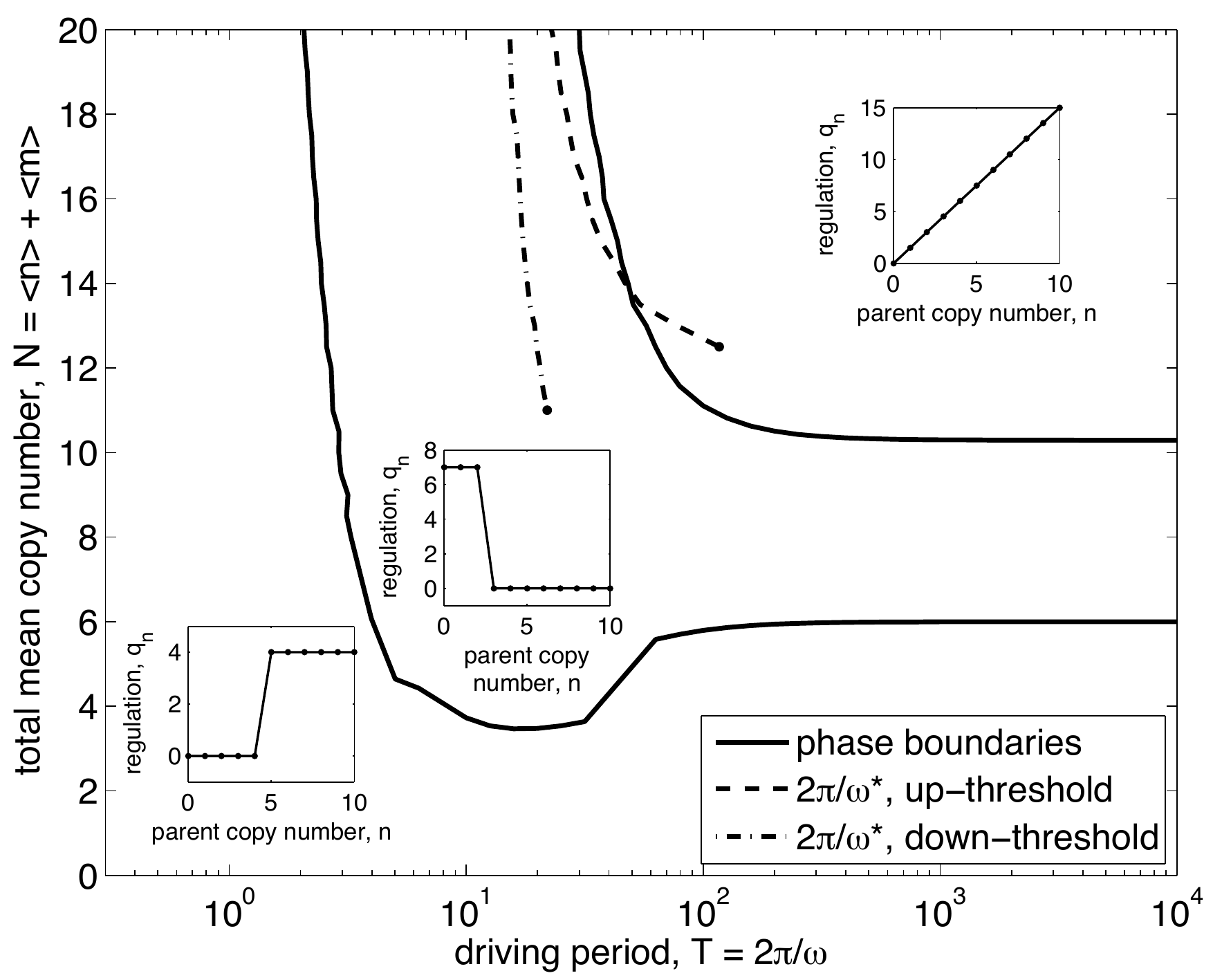}
\linespread{1}
\caption{Phase diagram showing best optimal information $I^*(\phi,m)$ among linear, up-threshold, and down-threshold regulation in the space of driving period and copy number.  Phases are separated by solid lines and marked by sample regulation functions (insets).  Also shown is  ($2\pi$ over) resonant frequency $\omega^*$ [see point c in Fig.\ \ref{omstar}(a)] as a function of copy number for both up- (dashed) and down-regulation (dot-dashed).}
\label{phase}
\end{center}
\end{figure}

At low copy number,
optimal information $I^*(\phi,m)$ behaves as one might expect from the small-oscillation limit (Eqn.\ \ref{Im}): it decreases monotonically with frequency [Fig.\ \ref{omstar}(a), bottom curves].  At high copy number,
$I^*(\phi,m)$ decreases monotonically with frequency for linear regulation but for threshold regulation exhibits a maximum at a resonant frequency [Fig.\ \ref{omstar}(a), top curves].
Careful examination of the child distribution
at different phases [Fig.\ \ref{omstar}(b-d)]
(or simply its mean,
Sec.\ \ref{mean})
reveals the origin of this maximum as follows.
As the parent oscillates about the threshold, the child distribution is switch-like, with two long-lived switch states centered at the threshold's low and high rates, and brief intermediate states in between.  At high copy number, the threshold rates are far apart (one is zero and the other is large), making the switch states well separated and transmitting (slightly more than, due to the intermediate states) the strict switch limit \cite{mtv} of $I^*(\phi,m) \sim 1$ bit.
For slow oscillations the intermediate states are symmetric [Fig.\ \ref{omstar}(b)], but for faster oscillations there is a
lag in transitioning from one switch state to the other,
making the intermediate states distinguishable
[Fig.\ \ref{omstar}(c)],
and transmitting
more information about phase.
Thus the resonant frequency $\omega^*$ balances the slowness required to avoid time-averaging [Fig.\ \ref{omstar}(d)] with the speed required for distinguishable intermediate states.  As seen in Fig.\ \ref{phase}, the onset of $\omega^*$ occurs above a critical copy number of $N^* \sim 10$.

Fig.\ \ref{omstar}(a) contains examples in which the most optimal regulation function is up-threshold (at low $N$), down-threshold (at high $N$ and high $\omega$) or linear (at high $N$ and low $\omega$).  The phase diagram (Fig.\ \ref{phase}) shows the results of this competition across a range of copy numbers $N$ and periods $T=2\pi/\omega$.
Linear regulation is best when both $N$ and $T$ are large, ultimately surpassing threshold regulation's
limit of $\sim$$1$ bit.  Down-threshold regulation is best at values of $T$ near $2\pi/\omega^*$ because its intermediate states are more distinguishable (i.e.\ have a larger Jensen-Shannon divergence) than those of a similarly parameterized up-threshold.  Up-threshold regulation is best at low $N$ due to its tendency, as discussed above (Eqn.\ \ref{gstar}) and in \cite{mtv}, to require fewer proteins to match the transmission across a similarly parameterized down-threshold.

The low-pass behavior revealed in Eqns. \ref{Ipar2} and \ref{Im} is consistent with our intuition from measuring static quantities in the presence of intrinsic noise \cite{berg}: the longer we wait, the more accurate is our estimate. However, in the presence of oscillatory driving, we find that threshold regulation can lead to an information-optimal frequency, and waiting longer is not necessarily the optimal strategy. Further, we have shown that, at a fixed allowed copy number and allowed integration time, one may find that a different regulation strategy (linear, threshold up-regulation, or threshold down-regulation) is optimal for responding to oscillatory driving. Absent from this analysis are intriguing questions such as whether the diversity of other network topologies observed in nature --- including cascades and feedback circuits --- are consistent with these observations. We anticipate the spectral method will continue to be useful in addressing these challenges.

\appendix

\section{The parent distribution is a Poisson with oscillating mean}
\label{parent}

In this study we model two transcription factors --- the `parent' with copy number $n$ and the `child' with copy number $m$ --- each undergoing a birth-death process, with the parent's birth rate an oscillatory function of time $f(t) = g+\alpha\cos\omega t$ and the child's birth rate an arbitrary function $q_n$ of the parent copy number.  The master equation,
\beqn
\label{ME}
\dot{p}_{nm} &=& f(t)p_{n-1,m}-f(t)p_{nm} \nonumber \\
&&	+(n+1)p_{n+1,m}-np_{nm} \nonumber \\
&&	+\rho[q_np_{n,m-1}-q_np_{nm} \nonumber \\
&&	+(m+1)p_{n,m+1}-mp_{nm}],
\eeqn
describes the time evolution of the joint probability of observing $n$ and $m$ copies of
the parent and child, respectively.  Here time is normalized via the parent decay rate (in these units $\rho$ is the child decay rate).

The equation for the parent distribution $p_n(t)$ is obtained by summing the master equation over $m$:
\beqn
\label{pME}
\dot{p}_n &=& f(t)p_{n-1}-f(t)p_n \nonumber \\
&&	+(n+1)p_{n+1}-np_n.
\eeqn
Since the parent is not regulated, Eqn.\ \ref{pME} simply describes a one-dimensional birth-death process with time-dependent birth rate $f(t)$.  The solution can be found, regardless of the form of $f(t)$, using either (i) the method of characteristics, or (ii) the spectral method; for completeness we present both.

\subsection{Method of characteristics}
\label{moc}

We begin the solution of Eqn.\ \ref{pME} by defining the generating function $G(x,t) = \sum_n p_n(t)x^n$ \cite{vanKampen} over complex variable $x$ (writing $x=e^{ik}$ makes clear that the generating function is the Fourier transform in copy number).  The utility of the generating function is that by summing Eqn.\ \ref{pME}, which describes an infinite set of ordinary differential equations in $p_n$, over $n$ against $x^n$, it becomes a single partial differential equation in $G$,
\beq
\label{Gx}
\dot{G} = -(x-1)[\partial_x-f(t)]G.
\eeq
The distribution is recovered by inverse transform, $p_n(t) = \partial^n_x\left[G(x,t)\right]_{x=0}/n!$.

We solve Eqn.\ \ref{Gx} by the method of characteristics, in which we demand that $x$ and $t$ lie along a characteristic line $(x(s), t(s))$ parameterized by $s$, and we seek $G(s)$.
Expressing Eqn.\ \ref{Gx} as
\beq
\label{Gx2}
(x-1)f(t)G = \frac{\partial G}{\partial t}+\frac{\partial G}{\partial x}(x-1),
\eeq
it is clear that its consistency with the chain rule
\beq
\frac{dG}{ds} = \frac{\partial G}{\partial t}\frac{dt}{ds}+\frac{\partial G}{\partial x}\frac{dx}{ds}
\eeq
requires
\beqn
\label{ts}
\frac{dt}{ds}=1 &\Rightarrow& t-t_0 = s-s_0, \\
\label{xs}
\frac{dx}{ds}=x-1 &\Rightarrow& y = y_0e^{s-s_0} = y_0e^{t-t_0}, \qquad \\
\label{Gs}
\frac{dG}{ds}=(x-1)f(t)G &\Rightarrow& \frac{dG}{dt} = y_0e^{t-t_0}f(t)G,
\eeqn
where $y\equiv x-1$, the last step of Eqn.\ \ref{xs} uses Eqn.\ \ref{ts}, and the last step of Eqn.\ \ref{Gs} uses Eqns.\ \ref{ts} and \ref{xs}.  We integrate Eqn.\ \ref{Gs},
\beq
\label{Gtemp}
G(y,t) = G_0\exp \left[y_0e^{-t_0} \int_{t_0}^t dt' \, e^{t'}f(t') \right],
\eeq
and recognize that the initial condition $G_0$ is an arbitrary function of $y_0$; we may therefore expand as  $G_0 = \sum_j c_j y_0^j$ for some $c_j$.  Inserting the characteristic equation $y_0 = ye^{-(t-t_0)}$ (Eqn.\ \ref{xs}), Eqn.\ \ref{Gtemp} becomes
\beq
\label{Gy}
G(y,t) = \sum_j c_j y^j e^{-j(t-t_0)}e^{\nu(t)y}
\rightarrow c_0 e^{\nu(t)y},
\eeq
where
\beq
\label{nudef}
\nu(t) \equiv e^{-t}\int_{t_0}^t dt' \, e^{t'}f(t'),
\eeq
and the last step in Eqn.\ \ref{Gy} takes $t_0 \rightarrow -\infty$ to give the post-transient behavior, upon which only the $j=0$ mode survives.  Inverse transforming we find 
\beq
\label{timePoisson}
p_n(t)
= \frac{1}{n!}\partial^n_x\left[ c_0 e^{\nu(t)(x-1)} \right]_0
= e^{-\nu(t)}\frac{\nu(t)^n}{n!}
\eeq
(where $c_0=1$ by normalization), a Poisson distribution with time-dependent mean $\nu(t)$.

For oscillatory driving $f(t) = g + \alpha\cos\omega t$, the mean evaluates to
\beqn
\nu(t) &=& g + \frac{\alpha}{1+\omega^2} \left( \cos \omega t + \omega \sin \omega t \right)\\
\label{nuapp}
	&=& \nu_0 + 2|\nu_1|\cos(\omega t-\gamma),
\eeqn
where $\nu_0 = g$, $2|\nu_1| = \alpha/\sqrt{1+\omega^2}$, and $\gamma = \tan^{-1}\omega$.  Eqn.\ \ref{nuapp} shows that the parent oscillates about the same point and with the same frequency as the driving birth rate, but that the oscillation is damped and phase-shifted at high frequency.

\subsection{Spectral method}
\label{sm}

Eqn.\ \ref{pME} can also be solved using the spectral method \cite{mtv,cspan}.  Again we employ the generating function, this time expanding in a state space $\ket{n}$ indexed by copy number $n$: $\ket{G(t)} = \sum_n p_n(t)\ket{n}$. (Projecting onto the position space $\bra{x}$ recovers the previous form with $\ip{x}{G(t)} = G(x,t)$ and $\ip{x}{n} = x^n$.)  Summing Eqn.\ \ref{pME} against $\ket{n}$ gives
\beq
\label{Gket}
\ket{\dot{G}} = -(\ad-1)[\al-f(t)]\ket{G} = -\bd\bl(t)\ket{G},
\eeq
where the operators $\ad$ and $\al$ raise and lower copy number, respectively, i.e.\ $\ad\ket{n} = \ket{n+1}$ and $\al\ket{n} = n\ket{n-1}$ \cite{doi,zeldovich,peliti,mattis}, and we define $\bd\equiv\ad-1$ and $\bl(t)\equiv\al-f(t)$.  (Note  $\ad \leftrightarrow x$ and $\al \leftrightarrow \partial_x$, as is clear from Eqns.\ \ref{Gket} and \ref{Gx}.)

The spectral method exploits the linearity of the master equation by expanding $\ket{G}$ in the eigenfunctions $\ket{j}$ of a
birth-death process \cite{mtv,cspan}.
Since here the birth rate $f(t)$ is time-dependent, we expand as
\beq
\ket{G} = \sum_jc_j\ket{j(t)}
\eeq
where the time-dependent functions $\ip{x}{j(t)} = (x-1)^je^{\nu(t)(x-1)}$ are parameterized by the (as yet unknown) function $\nu(t)$.  Noting that $\partial_t \ip{x}{j}
= \dot{\nu}\ip{x}{j+1}$, the left-hand side of Eqn.\ \ref{Gket} becomes $\ket{\dot{G}} = \sum_j\left[\dot{c}_j\ket{j} + \dot{\nu}c_j\ket{j+1}\right]$.  Defining $\bbar(t) \equiv \al -\nu(t) = \bl + f(t) -\nu(t)$ such that the $\ket{j(t)}$ are the eigenstates of $\bd\bbar$, i.e.\
\beq
\bd\bbar\ket{j(t)} = j\ket{j(t)}
\eeq
($\bd$ and $\bbar$ raise and lower $\ket{j}$ as $\ad$ and $\al$ do $\ket{n}$, respectively), the right-hand side of Eqn.\ \ref{Gket} becomes $\sum_j c_j \left\{-j\ket{j} + [f(t)-\nu(t)]\ket{j+1} \right\}$.  Therefore projecting $\bra{j'}$ onto Eqn.\ \ref{Gket} gives the following equation for the expansion coefficients $c_j$:
\beq
\label{cjeq}
\dot{c}_j + \dot{\nu}c_{j-1} = -jc_j + [f(t)-\nu(t)]c_{j-1}.
\eeq
The dynamics are trivial if $\dot{\nu} = -\nu + f(t)$, an equation whose solution is Eqn.\ \ref{nudef}.  In this case Eqn.\ \ref{cjeq} is solved by $c_j = e^{-j(t-t_0)}$, which becomes $\delta_{j0}$ as $t_0 \rightarrow -\infty$ (for post-transient behavior).  The probability distribution is obtained by inverse transform,
\beq
\label{timePoisson2}
p_n(t) = \ip{n}{G} = \sum_j c_j \ip{n}{j} = \ip{n}{0} = e^{-\nu(t)}\frac{\nu(t)^n}{n!},
\eeq
where the last step uses the fact the the zero mode $\ip{n}{0}$ is a Poisson distribution at the eigenfunction parameter (or gauge) $\nu(t)$ \cite{mtv,cspan}.  Eqn.\ \ref{timePoisson2} reproduces the result from the method of characteristics, Eqn.\ \ref{timePoisson}.

\section{Fourier transform of the parent distribution}
\label{ftp}

The Fourier coefficients of a Poisson distribution with oscillating mean are here found analytically in terms of spectral modes.  We begin by representing the distribution as
\beq
p(n|\phi) = e^{-\nu(\phi)}\frac{\nu(\phi)^n}{n!}
= \frac{1}{n!} \partial^n_x\left[ e^{\nu(\phi)(x-1)} \right]_0,
\eeq
where $\phi = \omega t$, and
\beq
\nu(\phi) = \nu_0 + 2|\nu_1|\cos(\phi-\gamma) = \nu_0 + |\nu_1|\sum_\pm e^{\pm i(\phi-\gamma)}.
\eeq
The Fourier transform will have support only at harmonics $z$ of the driving frequency, i.e.\
\beqn
p_n^z &=& \int_{0}^{2\pi} \frac{d\phi}{2\pi} e^{iz\phi} p(n|\phi) \\
	&=& \int_{0}^{2\pi} \frac{d\phi}{2\pi} e^{iz\phi} \frac{1}{n!}
	\partial^n_x \Biggl[ e^{\nu_0(x-1)} \nonumber \\
&&	\left. \times \exp \left\{ (x-1) |\nu_1| \sum_\pm e^{\pm i(\phi-\gamma)} \right\} \right]_0.
\eeqn
Expanding the exponential and then invoking the binomial expansion on the $\pm$ sum,
\beqn
p_n^z &=& \int_{0}^{2\pi} \frac{d\phi}{2\pi} e^{iz\phi} \frac{1}{n!}
	\partial^n_x \Biggl[ e^{\nu_0(x-1)} \nonumber \\
&&	\times \sum_{j=0}^\infty \frac{(x-1)^j|\nu_1|^j}{j!} \nonumber \\
&&	\left. \times \sum_{\ell=0}^j \frac{j!}{\ell!(j-\ell)!} e^{i(\phi-\gamma)(j-\ell)}
		e^{-i(\phi-\gamma)\ell} \right]_0.
\eeqn
Reordering terms,
\beqn
\label{pzbig}
p_n^z &=& \sum_{j=0}^\infty |\nu_1|^j \nonumber \\
&&	\times \frac{1}{n!} \partial^n_x \left[ e^{\nu_0(x-1)}(x-1)^j \right]_0 \nonumber \\
&&	\times \sum_{\ell=0}^j \frac{e^{-i(j-2\ell)\gamma}}{\ell!(j-\ell)!} \nonumber \\
&&	\times \int_{0}^{2\pi} \frac{d\phi}{2\pi} e^{i(j-2\ell+z)\phi}.
\eeqn
The second line in Eqn.\ \ref{pzbig} is the derivative representation of the spectral mode $\ip{n}{j}$ \cite{mtv,cspan},
with parameter (or gauge)
$\nu_0$.  The fourth line in Eqn.\ \ref{pzbig} evaluates to $\delta_{0,j-2\ell+z}$, which collapses the $\ell$ sum nonvanishingly provided $z+j$ is an even number from $0$ to $2j$, or $z \in \{-j,-j+2,\dots,j-2,j\}$.  This criterion can be equivalently be expressed as a condition on $j$ in terms of $z$: $j \in \{|z|,|z|+2,|z|+4,\dots,\infty\} \equiv \Omega$, making Eqn.\ \ref{pzbig}
\beq
\label{pnzsmall}
p_n^z = e^{iz\gamma} \sum_{j\in\Omega}
	\frac{|\nu_1|^j\ip{n}{j}}{[(j+z)/2]![(j-z)/2]!}
\eeq
Defining $j' \equiv (j-|z|)/2$ allows the sum to run from $0$ to $\infty$ and yields the simplification $[(j+z)/2]![(j-z)/2]! = j'!(j'+|z|)!$ for all integer $z$, making Eqn.\ \ref{pnzsmall}
\beq
\label{pnzsupp}
p_n^z = e^{iz\gamma} \sum_{j'=0}^\infty
	\frac{|\nu_1|^{2j'+|z|}}{j'!(j'+|z|)!}
	\left\langle n \big| 2j'+|z| \right\rangle,
\eeq
as in Eqn.\ 
\ref{pnz}.

\section{Expansion in the small-information limit}
\label{exp}

Here we explicitly expand the log in Eqn.\ 
\ref{IFourier}
and show that the first two terms contribute to the leading-order behavior in $p_n^z$.  Eqn.\ 
\ref{IFourier}
reads
\beq
\label{IFourierapp}
I(\phi,n) = \sum_{n,z}p_n^z\int_{0}^{2\pi} \frac{d\phi}{2\pi} e^{-iz\phi}
	\log \left(1+\sum_{z'\ne0}\frac{p_n^{z'}}{p_n^0}e^{-iz'\phi} \right).
\eeq
Since, for small $|\nu_1|$, Eqn.\ \ref{pnzsupp} is dominated by the $j'=0$ term, i.e.
\beq
\label{pnzapprox}
p_n^z\approx e^{iz\gamma} \frac{|\nu_1|^{|z|}}{|z|!} \left\langle n \big||z|\right\rangle,
\eeq
which is small for $z\ne0$, we expand the log:
\beqn
I &=& \sum_{n,z} p_n^z \int_0^{2\pi} \frac{d\phi}{2\pi} e^{-iz\phi} \nonumber \\
&&	\times\sum_{\ell=1}^\infty \frac{(-1)^{\ell+1}}{\ell} \left( \sum_{z'\ne 0}\frac{p_n^{z'}}{p_n^0}
	e^{-iz'\omega t} \right)^\ell,
\eeqn
or, explicitly writing out the $\ell$ sums,
\beqn
I &=& \sum_{n,z} p_n^z \int_0^{2\pi} \frac{d\phi}{2\pi} e^{-iz\phi}
	\sum_{\ell=1}^\infty \frac{(-1)^{\ell+1}}{\ell (p_n^0)^\ell} \nonumber \\
&&	\times \sum_{z_1\ne 0}p_n^{z_1} e^{-iz_1\phi}
	\sum_{z_2\ne 0}p_n^{z_2} e^{-iz_2\phi} \dots \nonumber \\
&&	\times \sum_{z_\ell\ne 0}p_n^{z_\ell} e^{-iz_\ell\phi}.
\eeqn
Reordering terms as
\beqn
I &=& \sum_{\ell=1}^\infty \frac{(-1)^{\ell+1}}{\ell} \sum_n \frac{1}{(p_n^0)^\ell} \nonumber \\
&&	\times \sum_{z_1\ne 0}p_n^{z_1}
	\sum_{z_2\ne 0}p_n^{z_2} \dots
	\sum_{z_\ell\ne 0}p_n^{z_\ell}
	\sum_z p_n^z \nonumber \\
&&	\times \int_0^{2\pi} \frac{d\phi}{2\pi}
	e^{-i(z+z_1 + z_2 + \dots + z_\ell)\phi},
\eeqn
and
employing $\int_{0}^{2\pi} d\phi\, e^{is\phi} = 2\pi\delta_{s0}$ for integer $s$
allows us to collapse the $z$ sum,
\beqn
\label{IL}
I &=& \sum_{\ell=1}^\infty \frac{(-1)^{\ell+1}}{\ell} \sum_n \frac{1}{(p_n^0)^\ell} \nonumber \\
&&	\times \sum_{z_1\ne 0}p_n^{z_1}
	\sum_{z_2\ne 0}p_n^{z_2} \dots
	\sum_{z_\ell\ne 0}p_n^{z_\ell} \nonumber \\
&&	\times p_n^{-(z_1 + z_2 + \dots + z_\ell)}.
\eeqn
Now, since $p_n^z \propto |\nu_1|^{|z|}$ for small $|\nu_1|$ (Eqn.\ \ref{pnzapprox}), $I(\phi,n)$ will be dominated by the leading-order term in $|z|$.  Writing out the first few terms in Eqn.\ \ref{IL} explicitly,
\beqn
I &=& \sum_n \frac{1}{p_n^0} \sum_{z\ne 0}p_n^z p_n^{-z} \nonumber \\
&&	- \frac{1}{2} \sum_n \frac{1}{(p_n^0)^2}
		\sum_{x\ne 0}p_n^x \sum_{y\ne 0}p_n^y p_n^{-(x+y)} \nonumber \\
&&	+ \frac{1}{3} \sum_n \frac{1}{(p_n^0)^3}
		\sum_{u\ne 0}p_n^u \sum_{v\ne 0}p_n^v \sum_{w\ne 0}p_n^w p_n^{-(u+v+w)} \nonumber \\
&&	+ \dots,
\eeqn
we see that the leading-order term is proportional to $p_n^1p_n^{-1} = |p_n^1|^2$ and has contributions from both the first ($z=\pm 1$) and the second ($(x,y)=(\pm 1,\mp 1)$) term in the log expansion.  To leading order, then,
\beq
\label{Ipar}
I \approx \left[1+1-\frac{1}{2}(1+1)\right] \sum_n \frac{|p_n^1|^2}{p_n^0}
	= \sum_n \frac{|p_n^1|^2}{p_n^0},
\eeq
as in Eqn.\ 
\ref{Ipar1}.

\section{Useful properties of the spectral modes}
\label{props}
\subsection{Relating spectral modes to distribution moments}
\label{m2m}

The eigenfunctions of a birth-death process, or `spectral modes,' define a complete basis in which any distribution $p_n$ can be expanded, i.e.\
\beq
p_n = \sum_j c_j \ip{n}{j}
\eeq
(where the expansion coefficients 
are computed by inverse transform: $c_j = \sum_n p_n \ip{j}{n}$).  The goal of this section is to relate the spectral modes to the distribution moments $\avg{n^\ell} = \sum_n n^\ell p_n$.

The $\ell$th moment is $\avg{n^\ell} = \sum_j c_j \eta^\ell_j$, where
\beq
\label{etaint}
\eta^\ell_j \equiv \sum_n n^\ell \ip{n}{j}
	= \sum_n n^\ell \oint \frac{dx}{2\pi i} \frac{(x-1)^j e^{g(x-1)}}{x^{n+1}}.
\eeq
The last step uses the integral representation of $\ip{n}{j}$ over complex variable $x$ \cite{cspan}.  Isolating the $n$-dependence, we may write
\beq
\label{eta1}
\eta^\ell_j 	= \oint \frac{dx}{2\pi i} (x-1)^j e^{g(x-1)}\frac{1}{x} S_\ell,
\eeq
where for $w \equiv 1/x$,
\beq
\label{Seq}
S_\ell \equiv \sum_n n^\ell w^n
	= \sum_n \left( w \partial_w \right)^\ell w^n
	= \left( w \partial_w \right)^\ell \frac{1}{1-w},
\eeq
and the last step sums the geometric series.  Computing the first few derivatives in Eqn.\ \ref{Seq} reveals the pattern
\beq
S_\ell = \frac{\sum_{u=0}^\ell A_{\ell u}w^{\ell-u}}{(1-w)^{\ell+1}}
	= \frac{\sum_{u=0}^\ell A_{\ell u}x^{u+1}}{(x-1)^{\ell+1}},
\eeq
in terms of the Eulerian numbers
\beq
A_{\ell u} = \sum_{v=0}^u (-1)^v
	\begin{pmatrix} \ell+1 \\ v \end{pmatrix}
	(u+1-v)^\ell,
\eeq
making Eqn.\ \ref{eta1}
\beq
\label{eta2}
\eta^\ell_j 	= \sum_{u=0}^\ell A_{\ell u} \oint \frac{dx}{2\pi i} \frac{x^u}{(x-1)^{\ell-j+1} e^{-g(x-1)}}.
\eeq
The integral is recognized as a representation of the conjugate mode $\ip{j'}{n'}$ with $j'=\ell-j \ge 0$, $n'=u \ge 0$, and parameter (or gauge) $-g$ \cite{cspan}.  Modes and conjugate modes can be evaluated either recursively  using selection rules or explicitly using Cauchy's theorem \cite{cspan}.  Conjugate modes $\ip{j'}{n'}$ with gauge $g'$ are $j'$th order polynomials in $n'$; the first few are
\beqn
\ip{j'=0}{n'} &=& 1, \\
\ip{j'=1}{n'} &=& n'-g', \\
\ip{j'=2}{n'} &=& \frac{1}{2} n'^2 - \left( g'+\frac{1}{2} \right) n' + \frac{g'^2}{2}.
\eeqn
Thus the first few $\eta^\ell_j$ are (Eqn.\ \ref{eta2})
\beqn
\label{eta3}
\sum_n \ip{n}{j} &=& \eta^{\ell=0}_j = \delta_{j0}, \\
\label{eta4}
\sum_n n \ip{n}{j} &=& \eta^{\ell=1}_j = g\delta_{j0} + \delta_{j1}, \\
\label{eta5}
\sum_n n^2 \ip{n}{j} &=& \eta^{\ell=2}_j \nonumber \\
	&=& g(g+1)\delta_{j0} + (2g+1)\delta_{j1} + 2\delta_{j2},
	\qquad
\eeqn
and in general the $\ell$th moment is calculated $\avg{n^\ell} = \sum_j c_j \eta^\ell_j$.

Eqns.\ \ref{eta3} and \ref{eta4} are useful in describing the mean properties of a child species regulated by the linear function $q_n = q_0 + cn$.  As described in Eqn.\ 
\ref{mu0},
the mean of the child distribution oscillates about the point
\beq
\mu_0 = \sum_n q_n p_n^0
	= \sum_n (q_0 + cn) \sum_j \frac{|\nu_1|^{2j}}{(j!)^2} \ip{n}{2j},
\eeq
where the last step uses the Fourier transform of the parent distribution, Eqn.\
\ref{pnzsupp}.  Evaluating the $n$ sum using Eqns.\ \ref{eta3} and \ref{eta4} gives
\beqn
\mu_0 &=& \sum_j \frac{|\nu_1|^{2j}}{(j!)^2} \left[ q_0 \delta_{2j,0} + c (g\delta_{2j,0} + \delta_{2j,1}) \right] \\
\label{mu0linear}
&=&	q_0 + cg,
\eeqn
(since only the first and second Kronecker delta have support for integer $j$), as in Eqn.\ 
\ref{mu0} (top).

Similarly, as described in Eqn.\ 
\ref{mu1},
the amplitude of the oscillation of the child mean is
\beqn
|\mu_1| &=& \frac{\sum_n q_n |p_n^1|}{\sqrt{1+(\omega/\rho)^2}}
	= \frac{1}{\sqrt{1+(\omega/\rho)^2}} \nonumber \\
&&	\times \sum_n (q_0 + cn) \sum_j \frac{|\nu_1|^{2j+1}}{j! (j+1)!} \ip{n}{2j+1}.
	\qquad
\eeqn
Again employing Eqns.\ \ref{eta3} and \ref{eta4},
\beqn
|\mu_1| &=& \frac{1}{\sqrt{1+(\omega/\rho)^2}} \sum_j \frac{|\nu_1|^{2j+1}}{j! (j+1)!} \nonumber \\
&&	\times \left[ q_0 \delta_{2j+1,0} + c (g\delta_{2j+1,0} + \delta_{2j+1,1}) \right] \\
\label{mu1linear}
&=&	\frac{c|\nu_1|}{\sqrt{1+(\omega/\rho)^2}}
\eeqn
(since only the last Kronecker delta has support for integer $j$), as in Eqn.\ 
\ref{mu1} (top).

\subsection{Sums of differences}
\label{ibp}

The zero mode (i.e.\ the steady-state solution) of the birth-death process with birth rate (or gauge) $g$ is the Poisson distribution,
\beq
\ip{n}{j=0} = e^{-g}\frac{g^n}{n!}
\eeq
(recall that time is normalized by the decay rate).  Each higher mode is related to the previous mode by discrete derivative \cite{cspan}:
\beq
\label{discrete}
\ip{n}{j+1} = -\partial_n^- \ip{n}{j} = \ip{n-1}{j} - \ip{n}{j}.
\eeq
This property is especially useful when performing a finite sum, since only the boundary terms survive:
\beq
\label{sbp}
\sum_{n=a}^b \ip{n}{j+1} = \ip{a-1}{j} - \ip{b}{j}.
\eeq

We make use of Eqn.\ \ref{sbp} in arriving at Eqns.\ 
\ref{mu0} (bottom) and \ref{mu1} (bottom),
describing the mean properties of a child species regulated by the threshold function $q_n = q_0+\Delta\chi(n\in\Omega_\pm)$.  Here $\chi$ is a characteristic function equal to $1$ when $n$ is in the set $\Omega_+=\{n>n_0\} = \{n_0+1, \dots, \infty\}$ (for up-regulation) or $\Omega_-=\{n\le n_0\} = \{0, \dots, n_0\}$ (for down-regulation), and $0$ otherwise.
In Eqn.\
\ref{mu0} (bottom),
\beqn
\mu_0 &=& \sum_n q_n p_n^0 \\
	&=& \sum_n [q_0 + \Delta\chi(n\in\Omega_\pm)] \sum_j \frac{|\nu_1|^{2j}}{(j!)^2} \ip{n}{2j}, \\
\label{mu0sbptemp}
	&=& q_0 + \Delta\sum_j \frac{|\nu_1|^{2j}}{(j!)^2} \sum_{n\in\Omega_\pm} \ip{n}{2j},
\eeqn
where the $q_0$ term collapses as in Eqn.\ \ref{mu0linear}.  The $j=0$ term reduces explicitly to $\Delta\pi_\pm$, where $\pi_\pm \equiv \sum_{n\in\Omega_\pm} \ip{n}{0}$.  For any $j'>0$ we may employ Eqn.\ \ref{sbp},
\beqn
\label{sbp1}
\sum_{n=n_0+1}^{\infty} \ip{n}{j'} &=& \ip{n_0}{j'-1}, \\
\label{sbp2}
\sum_{n=0}^{n_0} \ip{n}{j'} &=& -\ip{n_0}{j'-1},
\eeqn
(since the boundary terms at $n=-1$ and $\infty$ vanish), making Eqn.\ \ref{mu0sbptemp}
\beq
\mu_0 = q_0 + \Delta \left[ \pi_\pm \pm 
	\sum_{j>0} \frac{|\nu_1|^{2j}}{(j!)^2} \ip{n_0}{2j-1} \right],
\eeq
as in
Eqn.\ \ref{mu0} (bottom).
In Eqn.\
\ref{mu1} (bottom),
\beqn
|\mu_1| &=& \frac{\sum_n q_n |p_n^1|}{\sqrt{1+(\omega/\rho)^2}}\\
	&=& \frac{1}{\sqrt{1+(\omega/\rho)^2}}
	\sum_n [q_0 + \Delta\chi(n\in\Omega_\pm)] \nonumber \\
&&	\times \sum_j \frac{|\nu_1|^{2j+1}}{j! (j+1)!} \ip{n}{2j+1} \\
	&=&\frac{\Delta}{\sqrt{1+(\omega/\rho)^2}}
		\sum_j \frac{|\nu_1|^{2j+1}}{j! (j+1)!} \sum_{n\in\Omega_\pm}  \ip{n}{2j+1},
	\qquad
\eeqn
where the $q_0$ term vanishes as in Eqn.\ \ref{mu1linear}.  Again using Eqns.\ \ref{sbp1} and \ref{sbp2},
\beq
|\mu_1| = \Delta \sum_j \frac{|\nu_1|^{2j+1}}{j!(j+1)!} \ip{n_0}{2j},
\eeq
as in
Eqn.\ \ref{mu1} (bottom),
where for the down-threshold the negative sign is absorbed into the definition $\mu(t) \approx \mu_0 \pm 2|\mu_1|\cos(\omega t-\theta)$.

\begin{figure}
\begin{center}
\includegraphics[scale=0.71]{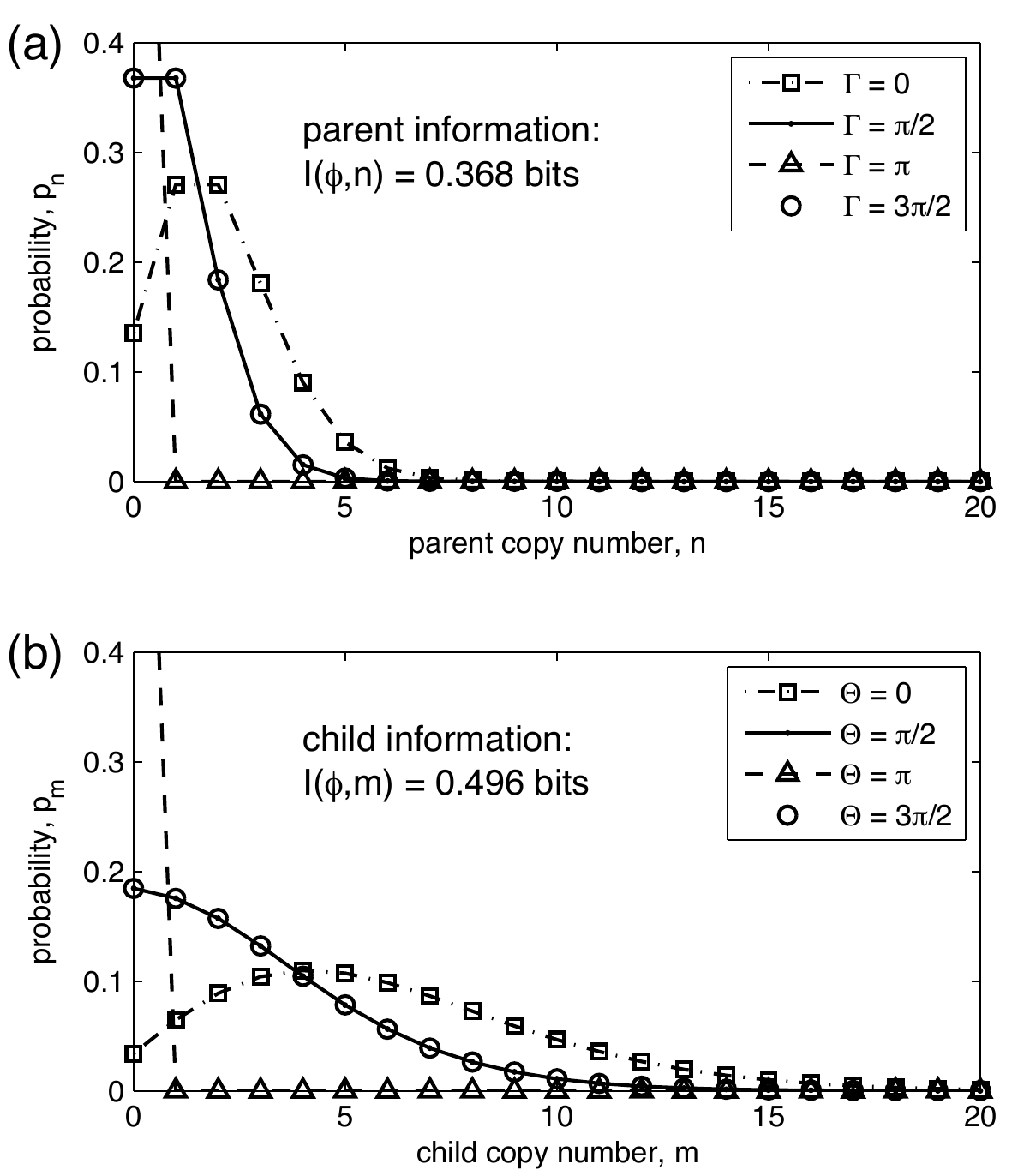}
\linespread{1}
\caption{The child can tell better time than the parent.  Parent (a) and child (b) distributions are plotted at phases $\Gamma \equiv \omega t - \gamma$ and $\Theta \equiv \omega t - \theta$ equal to $0$, $\pi/2$, $\pi$, and $3\pi/2$, where $\gamma$ and $\theta$ are the phase shifts of the parent and child means, respectively.  Regulation is linear and steep (slope $c=3$); other parameters are $g=\alpha=\rho=1$ and $\omega=q_0=0$.}
\label{childbetter}
\end{center}
\end{figure}

\begin{figure}
\begin{center}
\includegraphics[scale=1.1]{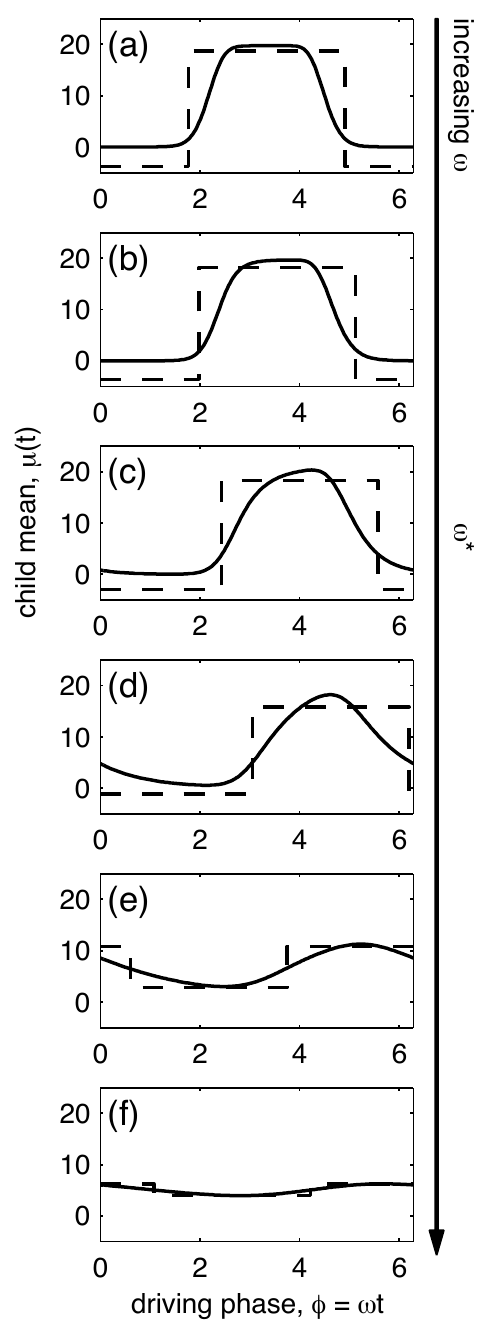}
\linespread{1}
\caption{Information-optimal child distribution mean $\mu(t) = \sum_m p_m(t)$ (solid) as a function of phase $\phi = \omega t$ for threshold down-regulation and fixed copy number $N = 20$ and driving frequency $\omega = $ (a) $0.10$, (b) $0.21$, (c) $0.46$, (d) $0.91$, (e) $1.91$, and (f) $3.98$.  Panel (c) corresponds to resonant frequency $\omega^*$ at which optimal information $I^*(\phi,m)$ is greatest.  For comparison, dashed lines show square waves centered at $\mu_0$ with amplitude $2|\mu_1|$ and phase shift $\theta = {\rm phase}(\mu_1)$.}
\label{sqwave}
\end{center}
\end{figure}

\subsection{Differentiation with respect to gauge}
The derivative of spectral mode $\ip{n}{j}$ with respect to its gauge $g$ is
\beq
\partial_g \ip{n}{j} = \ip{n}{j+1},
\eeq
a property that can be seen most readily from the integral representation of $\ip{n}{j}$ (Eqn.\ \ref{etaint}).  This property is useful in expediting derivative calculations, for example Eqn.\
\ref{gstar},
which uses
\beqn
\label{dn00}
\partial_g \ip{n_0}{0} &=& \ip{n_0}{1} = \ip{n_0}{0}\left( \frac{n_0}{g} - 1 \right), \\
\label{dpipm}
\partial_g \pi_\pm &=& \partial_g \sum_{n \in \Omega_\pm} \ip{n}{0} = \sum_{n \in \Omega_\pm} \ip{n}{1}
	= \pm \ip{n_0}{0},
	\qquad
\eeqn
where the last step in Eqn.\ \ref{dn00} uses Eqn.\ \ref{discrete}, and the last step in Eqn.\ \ref{dpipm} uses Eqns.\ \ref{sbp1} and \ref{sbp2}.

\section{The child can tell better time than the parent}
\label{nonmark}

Although the master equation (Eqn.\ \ref{ME}) is Markovian in time (i.e.\ the probability of making a transition at time $t$ is independent of previous transitions), it is not explicitly Markovian in the variables $t$, $n$, and $m$: $p(m|t)\ne\sum_np(m|n)p(n|t)$.  As such, information transmission is not bound by the data-processing inequality \cite{Cover}, and it is possible for the child to transmit more information than the parent about the driving phase, $I(\phi,m)>I(\phi,n)$.  This possibility is explicitly apparent in the small-oscillation limit with linear regulation,
Eqn.\
\ref{Im} (top),
\beq
\label{Imlinear}
I(\phi,m) = \frac{gc^2}{(q_0+cg)[1+(\omega/\rho)^2]}I(\phi,n),
\eeq
for example if $\omega\rightarrow0$, $q_0=0$, and $c>1$.  However, because these are the very parameter settings that strain the approximations under which Eqn.\ \ref{Imlinear} is derived (i.e.\ weak or fast oscillation and near-constant regulation), it is useful to also demonstrate numerically via the spectral method a case in which $I(\phi,m)>I(\phi,n)$.  Fig.\ \ref{childbetter} shows clearly that if the regulation is sufficiently steep, the oscillation is sufficiently amplified and the child tells better time than the parent does, i.e.\ $I(\phi,m)>I(\phi,n)$.

\section{Threshold regulation can exhibit a resonant frequency}
\label{mean}

When the regulation function is a threshold and the total copy number $N$ is sufficiently high,
the optimal information $I^*(\phi,m)$ exhibits a maximum at a resonant frequency $\omega^*$.
In Fig.\
\ref{omstar}
it is shown that $\omega^*$ is the frequency at which the driving oscillation is slow enough for the output to be switch-like (and avoid time-averaging), but fast enough for the brief states in between the switch states to be distinguishable from each other.

Fig.\ \ref{sqwave} here plots the mean $\mu(t)$ of the child distribution against phase for a range of driving frequencies. At low frequency [Fig.\ \ref{sqwave}(a)], the output is switch-like, and $\mu(t)$ is well approximated by a square wave.  At the resonant frequency [Fig.\ \ref{sqwave}(c)], the output is still switch-like but no longer symmetric in time, the asymmetry arising from a lag in transitioning from one switch state to the other (the lag allows the transitions to be distinguished from each other, maximizing transmission of information about phase).  At high frequency, [Fig.\ \ref{sqwave}(f)] the driving is faster than the parent decay rate, and both parent and child distributions are time-averaged.


\end{document}